\documentclass[twocolumn,showpacs,amsmath,aps,pra]{revtex4}

 \usepackage{graphicx}
 \usepackage{bm}

\begin{document}

 \title{Screened Casimir forces}
 \author{M. S. Toma\v s}
 \email{tomas@thphys.irb.hr}
 \affiliation{Rudjer Bo\v skovi\' c Institute, P. O. B. 180,
 10002 Zagreb, Croatia}
 \date{\today}

 \begin{abstract}
We demonstrate that a very recently obtained formula for the force
on a slab in a material planar cavity based on the calculation of
the vacuum Lorentz force [C. Raabe and D.-G. Welsch, Phys. Rev. A,
{\bf 71}, 013814 (2005)] describes a (medium) screened Casimir
force and, in addition to it, a medium-assisted force. The latter
force also describes the force on the cavity medium. For dilute
media, it implies the atom-mirror interaction of the
Casimir-Polder type at large and of the Coulomb type at small
atom-mirror distances of which the sign is insensitive to the
polarizability type (electric or magnetic) of the atom.

\end{abstract}
 \preprint{IRB-TH-1/05}
 \maketitle

It is well known that an atom in the vicinity of a body (mirror)
experiences the Casimir-Polder force \cite{CP} and, at smaller
distances, its nonretarded counterpart, the van der Waals force.
Consequently, being a collection of atoms, every piece of a medium
in front of a mirror should experience the corresponding force.
Despite this, a number of approaches to the Casimir effect
\cite{Cas} in material systems lead to the result that the Casimir
force on the medium between two mirrors vanishes and that the only
existing force is that between the mirrors \cite{Schw,Zhou,Tom02}
(see also text books \cite{Abr,Mil} and references therein). To
overcome this "unphysical" result, usually derived by calculating
the Minkowski stress tensor \cite{Schw,Tom02} but also obtained
using other methods \cite{Zhou,Mil}, Raabe and Welsch very
recently \cite{Raa04} suggested a Lorentz-force approach to the
Casimir effect. In their approach the force on a body is obtained
by calculating the sum of the vacuum Lorentz forces acting on its
constituents. Evidently, this method should lead to a nonzero
force on the medium between the mirrors. As an application of
their approach, Raabe and Welsch derived a formula for the force
on a magnetodielectric slab in a magnetodielectric planar cavity,
as depicted in Fig. 1. In this paper we i) demonstrate that,
according to the Raabe and Welsch formula, the total force on the
slab actually consists of a medium-screened Casimir force and a
medium-assisted force and ii) point out a few unexpected results
coming from the unusual properties of the latter force.

\begin{figure}[htb]
\label{sys}
 \begin{center}
 \resizebox{8cm}{!}{\includegraphics{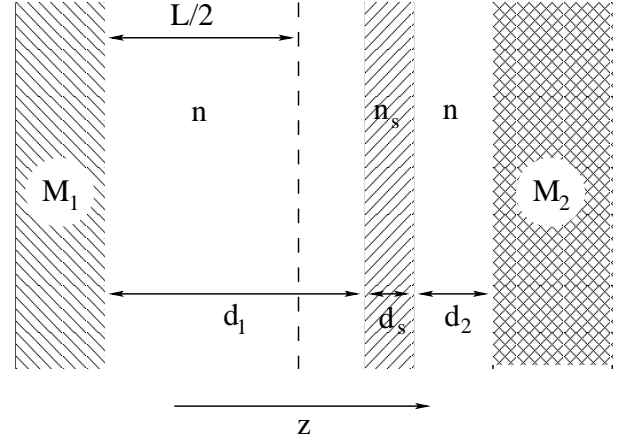}}
 \end{center}
 \caption{A slab in a planar cavity shown schematically. The
 (complex) refraction index of the slab is
 $n_s(\omega)=\sqrt{\varepsilon_s(\omega)\mu_s(\omega)}$ and that
 of the cavity $n(\omega)=\sqrt{\varepsilon(\omega)\mu(\omega)}$.
 The cavity walls are described by their reflection coefficients
 $r^q_1(\omega,k)$ and $r^q_2(\omega,k)$, with $k$ being the
 in-plane wave vector of a wave. The arrow indicates the direction
 of the force on the slab.}
\end{figure}

In the Lorentz-force approach, the force on the slab in the
configuration of Fig. 1 is given by \cite{Raa04}
\begin{eqnarray}
\label{fs} f(d_1,d_2)&=&-\frac{\hbar}{8\pi^2}\int_0^\infty d\xi
\int^\infty_0 \frac{dkk}{\kappa}\mu\times\nonumber\\
&&\sum_{q=p,s}\left[g_{q2}(i\xi,k;0)-g_{q1}(i\xi,k;d_1)\right],
\end{eqnarray}
where
\begin{equation}
 \kappa(\xi,k)=\sqrt{n^2(i\xi)\frac{\xi^2}{c^2}+k^2}
\end{equation}
is the perpendicular wave vector in the cavity at the imaginary
frequency, and \cite{com}
\[g_{q2}(i\xi,k;0)-g_{q1}(i\xi,k;d_1)
=-\left\{4\kappa^2\left(\delta_{qs}+\frac{1}{n^2}\delta_{qp}\right)r^q\right.\]
\begin{equation}
\label{g21}
\left.+\frac{\xi^2}{c^2}(n^2-1)[(1+r^q)^2-{t^q}^2]\Delta_q\right\}
\frac{r^q_2e^{-2\kappa d_{2}}-r^q_1e^{-2\kappa d_{1}}}{N^q},
\end{equation}
\begin{eqnarray}
N^q&=&1-r^q(r^q_1e^{-2\kappa d_{1}}+r^q_2e^{-2\kappa
d_2})\nonumber\\
&+&({r^q}^2-{t^q}^2)r^q_1r^q_2e^{-2\kappa (d_1+d_2)},
\end{eqnarray}
with $\Delta_q=\delta_{qp}-\delta_{qs}$. Here
$r^q=r^q_{1/2}=r^q_{2/1}$ and $t^q=t^q_{1/2}=t^q_{2/1}$ are
Fresnel coefficients for the (whole) slab given by
\begin{equation}
\label{rt} r^q(i\xi,k)=\rho^q\frac{1-e^{-2\kappa_s
d_s}}{1-{\rho^q}^2e^{-2\kappa_s d_s}},\;\; \;
t^q(i\xi,k)=\frac{(1-{\rho^q}^2)e^{-\kappa_s d_s}}
{1-{\rho^q}^2e^{-2\kappa_sd_s}},
\end{equation}
where
\begin{equation}
\rho^p(i\xi,k)=\frac{\varepsilon_s\kappa-\varepsilon\kappa_s}
{\varepsilon_s\kappa+\varepsilon\kappa_s},\;\;\;
\rho^s(i\xi,k)=\frac{\mu_s\kappa-\mu\kappa_s}
{\mu_s\kappa+\mu\kappa_s}, \label{rho}
\end{equation}
are the single-interface medium-slab ($\rho^q=r^q_{1s}=r^q_{2s}$)
Fresnel reflection coefficients.

\subsubsection{Medium-screened Casimir force and medium-assisted force}
Combining Eqs. (\ref{fs}) and (\ref{g21}), we see that $f$
naturally splits into two rather different components
\begin{equation}
\label{fsf} f(d_1,d_2)=f^{(1)}(d_1,d_2)+f^{(2)}(d_1,d_2),
\end{equation}
where
\[f^{(1)}(d_1,d_2)=\frac{\hbar}{2\pi^2}\int_0^\infty
d\xi \int^\infty_0dkk\kappa\times\]
\begin{equation}
\label{f1}
\sum_{q=p,s}\left(\mu\delta_{qs}+\frac{1}{\varepsilon}\delta_{qp}\right)
r^q\frac{r^q_2e^{-2\kappa d_{2}}-r^q_1e^{-2\kappa d_{1}}}{N^q},
\end{equation}
and
\[f^{(2)}(d_1,d_2)=\frac{\hbar}{8\pi^2c^2}\int_0^\infty
 d\xi\xi^2\mu(n^2-1)\int^\infty_0\frac{dkk}{\kappa}\times\]
\begin{equation}
\label{f2}
 \sum_{q=p,s}[(1+r^q)^2-{t^q}^2]\Delta_q
\frac{r^q_2e^{-2\kappa d_{2}}-r^q_1e^{-2\kappa d_{1}}}{N^q}.
 \end{equation}
Equation (\ref{f1}) differs in two respects from the formula for
the Casimir force in a dielectric cavity obtained through the
Minkowski tensor calculation \cite{Tom02}. First, the Fresnel
coefficients refer here to a magnetodielectric system. Another new
feature in Eq. (\ref{f1}) is the (effective) screening of the
force through the multiplication of the contributions coming from
TE- and TM-polarized waves by $\mu$ and $1/\varepsilon$,
respectively. This provides a simple recipe how to adapt the
traditionally obtained formulas for the Casimir force to the
present approach.

Clearly, $f^{(2)}$ owes its appearance to the cavity medium, note
that it vanishes when $n=1$, and is therefore a genuine
consequence of the Lorentz-force approach. Another unique feature
of $f^{(2)}$ is the dependence on the properties of the cavity
mirrors coming from its proportionality to $\Delta_q r^q_i$ rather
than to $r^q r^q_i$, as is the case with $f^{(1)}$. Owing to this
property the sign of each term in Eq. (\ref{f2}) depends on
whether the corresponding mirror is dominantly conductive
(dielectric) or permeable irrespective of the properties of the
slab. We illustrate this by calculating the force on an ideally
reflecting slab in a semi-infinite, e.g., $d_1\rightarrow\infty$,
cavity with an ideally reflecting mirror. Letting $t^q=0$,
$r^q=\pm\Delta_q$ [the minus sign is for an infinitely permeable
slab, see Eqs. (\ref{rt}) and (\ref{rho})], $r^q_2=\pm\Delta_q$,
and assuming $d_2\equiv d$ so large that $\varepsilon$ and $\mu$
can be replaced by their static values $\varepsilon_0$ and
$\mu_0$, respectively, the integrals in Eqs. (\ref{f1}) and
(\ref{f2}) become elementary and we find
\begin{equation}
f^{(1)}_{\rm id}(d)=\left\{\begin{array}{c}
1\\-\frac{7}{8}\end{array}\right\} \frac{\hbar c\pi^2}{15\cdot
2^5d^4}\sqrt{\frac{\mu_0}{\varepsilon_0}}
\left(1+\frac{1}{n_0^2}\right), \label{f1id}
\end{equation}
\begin{equation}
f^{(2)}_{\rm id}(d)=\pm\left\{\begin{array}{c}
1\\\frac{7}{8}\end{array}\right\} \frac{\hbar c\pi^2}{45\cdot
2^5d^4}\sqrt{\frac{\mu_0}{\varepsilon_0}}
\left(1-\frac{1}{n_0^2}\right).\label{f2id}
\end{equation}
Here, the first (second) line in the curly brackets corresponds to
a system with the slab and the mirror of the same (different
\cite{Boy}) type (conductive or permeable), whereas the sign of
$f^{(2)}_{\rm id}$ depends on whether the mirror is conductive (+)
or permeable (-). The above equations therefore describe the force
on the slab in four possible different configurations. Thus, the
result quoted by Raabe and Welsch \cite{Raa04} (the second line is
for optically dense cavity media)
\begin{equation}
f^{\rm cc}_{\rm id}(d)=\frac{\hbar
c\pi^2}{720d^4}\sqrt{\frac{\mu_0}{\varepsilon_0}}
\left(2+\frac{1}{n_0^2}\right)\simeq \frac{\hbar
c\pi^2}{360d^4}\sqrt{\frac{\mu_0}{\varepsilon_0}}, \label{fid}
\end{equation}
is recovered when the mirror and the slab are both conductive
(cc). Note that in this case and for dense media, $f^{(2)}_{\rm
id}$ is only three times smaller than $f^{(1)}_{\rm id}$. We also
observe that, when the mirror and the slab are both permeable
(pp), the above equations imply the total force $f^{\rm pp}_{\rm
id}=[(n^2_0+2)/(2n^2_0+1)]f^{\rm cc}_{\rm id}$ ($\simeq 1/2f^{\rm
cc}_{\rm id}$ for dense media). When the mirror and the slab are
of different type (cp or pc), however, we find $f^{\rm cp}_{\rm
id}=-(7/8)[(n^2_0+2)/(2n^2_0+1)]f^{\rm cc}_{\rm id}$ whereas
$f^{\rm pc}_{\rm id}=-(7/8)f^{\rm cc}_{\rm id}$. Finally, we note
that the force on the slab in a finite cavity is given by $f_{\rm
id}(d_1,d_2)=f_{\rm id}(d_2)-f_{\rm id}(d_1)$ and is therefore
obtained by combining the above results independently for each
part of the cavity.

\subsubsection{Force on the cavity medium and on an atom}
Specially, in the case $n_s=n$, $f=f^{(2)}$ describes the force on
a layer of the medium in the cavity. Letting $n_s=n$ [$\rho^q=0$
in Eq. (\ref{rt})], we have $r^q=0$ and $t^q=e^{i\beta d_s}$ in
Eq. (\ref{f2}), so that we may write
\begin{equation}
\label{fmz} f^{(2)}(d_1,d_2)=\int^{d_2+d_s}_{d_2}f_2(z)dz
-\int^{d_1+d_s}_{d_1}f_1(z)dz,
\end{equation}
where the force densities $f_i(z)$ are given by
\begin{eqnarray}
\label{fiz} f_i(z)&=&\frac{\hbar}{4\pi^2c^2}\int_0^\infty
d\xi\xi^2\mu (n^2-1)\int^\infty_0dkke^{-2\kappa z}\times
\nonumber\\
&&\sum_{q=p,s}\Delta_q \frac{r^q_i}{1-r^q_1r^q_2e^{-2\kappa L}},
\end{eqnarray}
with $L=d_1+d_2+d_s$ being the cavity length. For a semi-infinite
cavity (obtained by letting either $d_1\rightarrow\infty$ or
$d_2\rightarrow\infty$), $f_i(z)$ become characteristic functions
only of the medium and the corresponding mirror, so that we may
drop the index $i$ denoting the mirror. As follows from the above
definition, positive $f(z)$ means attraction between the medium
and the mirror. Clearly, for a dilute medium, the force density is
related to the force on an atom $f_{\rm at}(z)$ through
\begin{equation}
f(z)=Nf_{\rm at}(z),
\end{equation}
where $N$ is the atomic number density. The behavior of $f(z)$ is
therefore the same as that of $f_{\rm at}(z)$ discussed below.

Assuming the medium dilute and letting
\begin{equation}
n^2(i\xi)-1\simeq 4\pi N \alpha(i\xi),\;\;\;
\alpha(i\xi)=\alpha_e(i\xi)+\alpha_m(i\xi), \label{dil}
\end{equation}
where $\alpha_{e(m)}$ is the electric (magnetic) polarizability of
the atom, from Eq. (\ref{fiz}) (with $L\rightarrow\infty$) we find
\begin{eqnarray}
\label{faz} f_{\rm at}(z)&=&\frac{\hbar}{\pi c^2}\int_0^\infty
d\xi\xi^2 \mu\alpha\int^\infty_0dkke^{-2\kappa z}\times\nonumber\\
&&\left[r^p(i\xi,k)-r^s(i\xi,k)\right].
\end{eqnarray}
The integral over $\xi$ here effectively extends up to a frequency
$\Omega$ beyond which the mirror becomes transparent. Therefore,
at small atom-mirror distances $\Omega z/c\ll 1$, the main
contribution to the integral comes from large $k$'s ($k\sim 1/z$).
In this $k$-region, we may approximate the integrand with its
nonretarded (nr) counterpart obtained formally by letting
$\kappa=k$ and $\kappa_l=k$ for all layers of the mirror. In this
way, and by making the substitution $u=2kz$, we obtain
\begin{eqnarray}
\label{fanr} f_{\rm at}(z)&=&\frac{\hbar}{4\pi
c^2z^2}\int_0^\infty d\xi\xi^2\mu\alpha\int^\infty_0duue^{-u}
\times \nonumber\\
&&\left[r^p_{\rm nr}(i\xi,\frac{u}{2z})- r^s_{\rm
nr}(i\xi,\frac{u}{2z})\right].
\end{eqnarray}
For a single-medium mirror, $r^q_{\rm nr}(i\xi,u/2z)$ are
independent of $u$ [see Eq. (\ref{rho}), with $k\rightarrow\infty$
and $\{\varepsilon_s,\mu_s\}\rightarrow\{\varepsilon_m,\mu_m\}$]
and for this classical configuration we find
\begin{equation}
\label{fasm} f_{\rm at}(z)=\frac{\hbar}{4\pi c^2z^2}\int_0^\infty
d\xi\xi^2 \mu\alpha\left[\frac{\varepsilon_m-\varepsilon}
{\varepsilon_m+\varepsilon}-\frac{\mu_m-\mu}{\mu_m+\mu}\right]
\end{equation}
rather than the common van der Waals force.

To find the large-$z$ behavior of $f_{\rm at}(z)$, we make the
standard substitution $\kappa=n\xi p/c$ in Eq. (\ref{faz}). This
gives
\begin{eqnarray}
 f_{\rm at}(z)&=&\frac{\hbar}{\pi c^4}\int_0^\infty d\xi\xi^4
 \mu n^2\alpha\int^\infty_1dppe^{-2n\xi pz/c}\times\nonumber\\
 &&\left[r^p(i\xi,p)-r^s(i\xi,p)\right], \label{faz2}
\end{eqnarray}
where $r^q(i\xi,p)$ are obtained from $r^q(i\xi,k)$ by letting
$\kappa_l\rightarrow n(\xi/c)s_l$, with
$s_l=\sqrt{p^2-1+n^2_l/n^2}$ for all relevant layers. Thus, for
example, for a single-medium mirror with the refraction index
$n_m$ we have
\begin{equation}
\label{rps} r^p(i\xi,p)=\frac{\varepsilon_m p-\varepsilon s_m}
{\varepsilon_m p+\varepsilon s_m},\;\;\; r^s(i\xi,p)=\frac{\mu_m
p-\mu s_m} {\mu_m p+\mu s_m}.
\end{equation}
Now, for large $z$, the contributions from the region $\xi\simeq
0$ dominate the integral in Eq. (\ref{faz2}) and we may
approximate the frequency-dependent quantities with their static
values (denoted by the subscript $0$). In this case, the integral
over $\xi$ becomes elementary and we find
\begin{equation}
\label{flz} f_{\rm at}(z)=\frac{3\hbar c\alpha_0}{4\pi
n_0\varepsilon_0z^5}\int^\infty_1\frac{dp}{p^4}
\left[r^p(0,p)-r^s(0,p)\right].
\end{equation}
For a perfectly conductive mirror, the value of the above integral
is 2/3. As seen, since $n_0\varepsilon_0\simeq 1$ for dilute
media, in this case $f_{\rm at}(z)$ at large distances is
effectively three times smaller than the Casimir-Polder force
\cite{CP}. However, contrary to the behavior of the Casimir-Polder
force \cite{Boy2}, for an atom near a dominantly conductive
(permeable) mirror Eq. (\ref{flz}) [as well as Eqs.
(\ref{faz})-(\ref{fasm})] predicts an attractive (repulsive) force
irrespective of the polarizability of the atom \cite{com2}.

We end this short discussion by noting that ten years ago Zhou and
Spruch (ZS) considered the atom-mirror interaction for an atom in
a dielectric cavity \cite{Zhou}. Their result for the force on an
atom in a semi-infinite cavity is given by Eq. (\ref{faz}) when
letting
\begin{equation}
r^p(i\xi,k)\rightarrow
\left(2\frac{\kappa^2c^2}{n^2\xi^2}-1\right)r^p(i\xi,k)
\end{equation}
in the last factor of the integrand. Proceeding as before, we see
that for the leading term at small distances, this formula gives
the van der Waals force
\begin{equation}
\label{fvdW} f^{\rm ZS}_{\rm at}(z)=\frac{\hbar}{8\pi
z^4}\int_0^\infty
d\xi\frac{\alpha}{\varepsilon}\int^\infty_0duu^3e^{-u} r^p_{\rm
nr}(i\xi,\frac{u}{2z}).
\end{equation}
At large distances, it leads to
\begin{equation}
\label{fCP} f^{\rm ZS}_{\rm at}(z)=\frac{3\hbar c\alpha_0}{4\pi
n_0\varepsilon_0z^5}\int^\infty_1\frac{dp}{p^4}
\left[(2p^2-1)r^p(0,p)-r^s(0,p)\right],
\end{equation}
thus reproducing the Casimir-Polder result \cite{CP} in the case
of a perfectly conductive mirror and an empty cavity. Of course,
one must keep in mind that the physical situation considered by
Zhou and Spruch is substantially different from that considered in
this work. They calculated the force on an atom embedded in the
medium and not the force on an atom of the medium, as we have
done. Thus, although we may formally let $n_0\varepsilon_0\simeq
1$ in Eq. (\ref{flz}), as appropriate for a dilute medium, we
cannot interpret this as effectively the force on the atom in
vacuum. In other words, contrary to Zhou and Spruch, we cannot, in
principle, reproduce the Casimir-Polder result starting from the
medium-assisted force and, for this reason, we may regard Eq.
(\ref{flz}) as describing a (medium) screened Casimir-Polder
force.

In conclusion, the Raabe and Welsch result for the force on a slab
in a planar cavity naturally splits into a formula for a
medium-screened Casimir force and into a formula for a
medium-assisted force. The latter force is in an unusual way
related to the properties of the cavity medium and mirrors. It
also describes the force on the cavity medium and, for dilute
media, implies the atom-mirror interaction of the screened
Casimir-Polder type at large and of the Coulomb type at small
atom-mirror distances. Contrary to the sign of the Casimir-Polder
interaction, the sign of the medium-assisted interaction is
insensitive to the polarizability type of the atom. Evidently, to
understand these results, a microscopic consideration of the
atom-mirror interaction for an atom of the medium in the vicinity
of a mirror is needed.

This work was supported by the Ministry of Science and Technology
of the Republic of Croatia under contract No. 00980101.


\begin{thebibliography}{00}
\bibitem{com2}Note that the sign of $f_{\rm at}$ would be consistent with the
sign of the Casimir-Polder force provided that
$\alpha=\alpha_e-\alpha_m$. This demands, however, that
$\varepsilon\simeq 1+4\pi\alpha_e$ but $\mu\simeq 1-4\pi\alpha_m$
in Eq. (\ref{dil}).
\end{thebibliography}
 \end{document}